\documentclass{kapproc} 

\usepackage{procps} 

\usepackage[dvips]{graphicx}

\upperandlowercase

\kluwerbib 

\begin{document}

\articletitle[]{Peering Through the Muck:  
Notes on the Influence of the Galactic 
Interstellar Medium on Extragalactic Observations}


\author{Felix J. Lockman}
\affil{National Radio Astronomy Observatory\footnote{The National Radio 
Astronomy Observatory is operated by Associated Universities, Inc., 
under a cooperative agreement with the National Science Foundation.} \\ 
Green Bank, WV USA}
\email{jlockman@nrao.edu}

\begin{abstract} 
This paper considers some effects of foreground  Galactic gas on 
 radiation received from 
 extragalactic objects, with an  emphasis on the use of the 21cm line 
to determine the total $N_{HI}$. In general, the opacity of the 21cm line 
makes it impossible to derive an accurate value of $N_{HI}$
by simply applying a formula to the observed emission, 
except in directions where there is very little interstellar matter. 
The 21cm line can be used to estimate the likelihood that there is significant 
 $H_2$ in a particular direction, but carries little or no information on 
the amount of ionized gas, which can be a major source of foreground effects. 
Considerable discussion is devoted to the importance of 
small-scale angular structure in HI, with the 
conclusion that it will rarely contribute significantly to the total  
 error compared to other factors (such as the  effects of 
ionized gas) for extragalactic sight lines  at high Galactic latitude.
 The direction of the Hubble/Chandra Deep Field North is used as an example of 
the complexities that might occur even in the absence of opacity or 
molecular gas. 

\end{abstract}


\section{Introduction}
The Interstellar Medium (ISM) regulates the evolution of the Galaxy.  
It is the source of 
material for new stars and the repository of the products of stellar 
evolution.    But it is 
a damned nuisance to astronomers seeking to peer beyond the local gas. 
 In this article I treat the ISM as if it were simply an impediment to 
knowledge, and suggest ways that one might estimate its effects. 
This topic has taken on increasing importance in recent years as 
more and more experiments are requiring correction for 
 the ``Galactic foreground'' (e.g., Hauser 2001).  Here the 
 emphasis will be on the use of the 21cm line 
to determine a total $N_{HI}$, for the 21cm line 
 is our most general tool, and $N_{HI}$ is an 
important quantity which can be used 
to estimate  $N_{He}$, E(B--V)  and $S_{100\mu}$, as well as the  likelihood 
that there is molecular hydrogen along the line of sight. 
Some of the points treated here are discussed in more detail in 
reviews by Kulkarni \& Heiles (1987),  
Dickey \& Lockman (1990; hereafter DL90), Dickey (2002), and 
 Lockman (2002).

\section{General Considerations}

Figure 1 shows the amount of neutral interstellar gas, expressed as an 
equivalent HI column density, needed to produce 
unity opacity given normal abundances.  
Below the C-band edge at 0.25 KeV the opacity results almost
entirely from photoelectric absorption by  hydrogen and  helium, which 
contribute about equally to $\tau$ (Balucinska-Church \& 
McCammon, 1992).

\begin{figure}
{\includegraphics[height=0.9\hsize,width=3in,angle=-90]{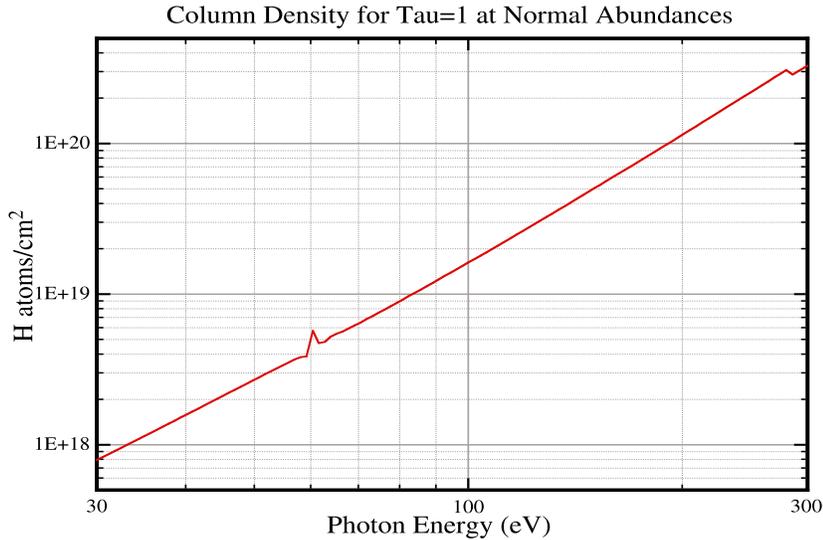}}
\caption{The equivalent $N_{HI}$ needed to produce an opacity of 
unity for the Galactic ISM at normal abundances, as a function of photon energy 
(Balucinska-Church \& McCammon 1992).
}
\end{figure}

Surveys of the sky in the 21cm line find 
 $\langle N_{HI} sin|b|\rangle  = 3 \times 10^{20}$ cm$^{-2}$ 
 where b is the Galactic latitude (DL90).   
Thus, {\it most}  sight-lines through the Milky Way  have $\tau \geq 1$ for 
$13.6 <  E < 300 $ eV, even without taking into account any contribution 
to the opacity from molecular hydrogen, $H_2$.    
Luckily, there are large areas of the sky over which $N_{HI}$ is 
a factor of several below the average, but even so,  the lowest 
$N_{HI}$ in {\it any} known direction is $4.4 \times 10^{19}$ cm$^{-2}$ 
(Lockman, Jahoda \& McCammon 1986; Jahoda, Lockman \& McCammon 1990), 
so  observations at $13.6 < E \leq  100$ 
 eV must always involve substantial corrections for the Galactic ISM. 

Except in directions where it is possible to make 
a direct measurement of $N_{HI}$ in  UV  absorption lines, 
every attempt to determine the effect of the ISM on extragalactic observations
should begin with the 21cm line.  
The Leiden-Dwingeloo (LD) 21cm survey covering $\delta > -30^{\circ}$ 
 at $35'$ angular resolution (Hartmann \& Burton 1997) 
supersedes all previous general 
surveys because of its angular and velocity resolution,
 and high quality of data.  A southern extension will be completed soon.  
Some parts of the sky of special interest 
have been mapped at higher angular resolution (e.g., Elvis et al. 1994; 
Miville-Desch\^{e}nes et al. 2002; 
Barnes \& Nulsen 2003).  The brighter emission near the 
Galactic plane is now being  measured at $1'$ resolution 
by a consortium who employ three different synthesis arrays 
(Knee 2002; McClure-Griffiths 2002; Taylor et al. 2002). 

\section{Estimating $N_{HI}$ from  21cm HI Data}

Radio telescopes measure an HI brightness temperature, $T_b$, as 
a function of velocity, but 
in general, there is no single formula that can be applied 
to derive   $N_{HI}$ from the observed  $T_b$. 
The solution to the equation of transfer for 21cm emission 
from a uniform medium is simple enough: 
\begin{equation}
T_b(v) = T_s[1-exp(-\tau(v))], 
\end{equation}
where $\tau(v) = 5.2 \times 10^{-19} N_{HI} / (T_s\ \Delta v)$ 
 for a Gaussian profile from a 
uniform cloud of linewidth  $\Delta v$ (FWHM) in km s$^{-1}$. 
$T_s$ is the excitation temperature of the transition, 
which is often, but not always, equal to the 
gas kinetic temperature (e.g., Liszt 2001).  But the real interstellar 
medium is not uniform, and the typical 21cm profile 
consists of several blended components formed in 
regions of different temperature.  If 
the line is optically thin  at all velocities 
there is no dependence of $N_{HI}$ on $T_s$ 
and $N_{HI} = 1.8 \times 10^{18} \int T_b dv$ cm$^{-2}$.  
The optically thin assumption always gives a lower limit on  $N_{HI}$.  

In directions where  part of the line has $\tau \geq 0.1$  
 the concept of a meaningful $T_s$ becomes ambiguous and 
 there is no unique solution for $N_{HI}$ from 21cm emission  data 
alone (e.g., Kalberla et al. 1985; DL90; Dickey 2002).
An HI cloud at 100 K with $\Delta v = 10$ km s$^{-1}$ has 
 $\tau = 0.1$ for $N_{HI} = 2 \times 10^{20}$ cm$^{-2}$, so 
the 21cm line in an average direction (see Fig.~2) 
should be treated as if it has components which are not optically thin.  

\subsubsection{Digression: The Two-phase ISM}

Theory tells us that under some conditions  HI can exist in 
 two stable phases at a single pressure:  a warm phase whose 
temperature is thousands of Kelvins, and a cool phase whose temperature is
$\leq100$ K (e.g., Field, Goldsmith, \& Habing 1969; 
Wolfire et al. 2003).  Observations suggest that reality is not so 
bimodal (e.g., Liszt 1983), but the 
generalization is still useful --- the ISM does contain 
cool HI with a high 21cm line opacity and warm HI with a low opacity 
(e.g., Heiles \& Troland 2003).   In the Solar neighborhood there is 
more  mass in the warm HI than the cold (Liszt, 1983; Dickey \& Brinks, 1993). 
 The cold phase fills a much smaller volume than the warm phase 
 and has a smaller scale-height as well, so at high 
Galactic latitudes many sight lines skirt the 
 clouds and intersect predominantly   ``intercloud'' medium, which 
has a low opacity because of its high temperature and turbulence.
In these directions $N_{HI}$ can be determined quite well.

\begin{figure}
{\includegraphics[height=0.6\vsize,width=3.0in,angle=-90]{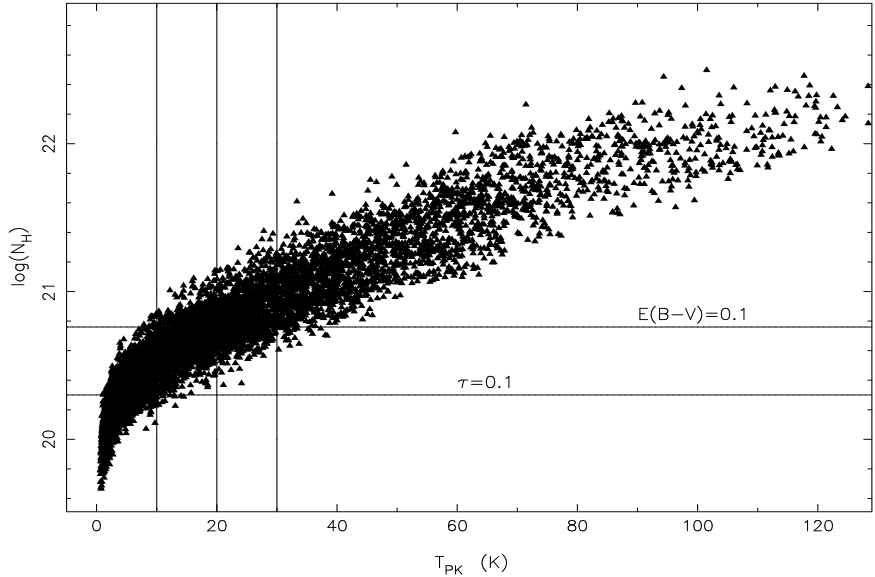}}
\caption{
The total $N_{HI}$ calculated for $T_s = 135$ K vs the peak line brightness
 temperature,  Tpk, 
for about $10^4$ 21cm spectra from the LD survey.  The sample 
includes data from $-90^{\circ} \leq b \leq 90^{\circ}$ 
 every $5^{\circ}$ in longitude between $30^{\circ}$ and 
$180^{\circ}$. Spectra above the $\tau = 0.1$ line likely contain 
components which have some opacity.  Directions with HI spectra 
above the $E(B-V)=0.1 $ line likely intersect regions of some 
molecular gas.
}
\end{figure}

\paragraph{Deriving $N_{HI}$ In Practice}
A 21cm profile should always be evaluated 
for  several excitation temperatures, 
 e.g.,  $T_s = 1000$ K and $135$ K.  If the resulting  
values of $N_{HI}$ differ significantly, 
where ``significantly'' depends on the accuracy one needs,  it 
is likely that a thorough investigation of the direction of interest 
must be made by measuring $\tau(v)$ in absorption  against nearby 
radio continuum sources ( DL90; Dickey 2002).  
A sense of how often this might be required is
given in Figure 2, where  $N_{HI}$  is 
plotted against the peak $T_b$ in the line, 
$T_{pk}$,  for a sample of $10^4$ spectra from the LD survey. 
The line  at $2 \times 10^{20}$ cm$^{-2}$ labeled $\tau = 0.1$ marks,
very approximately, the $N_{HI}$ below which 
the 21cm line is most likely optically thin.  
Conversely, Fig.~2 suggests that any  profile with $T_{pk} > 10$ K 
  should be tested for possible opacity. 
This applies to about $80\%$ of the directions in the Figure. 

\section{Angular Structure}

\begin{quote}
 The Mississippi [river] is remarkable in still another way --- its 
disposition to make prodigious jumps by cutting through 
narrow necks of land, and thus straightening and shortening itself...
In the space of 176 years the Lower Mississippi has 
shortened itself 242 miles.  That is an average of a trifle over 
one mile and a third per year.  Therefore, any calm person, who is 
not blind or idiotic, can see that in the Old Oolitic Silurian Period, 
just a million years ago next November, the Lower Mississippi River 
was upwards of 1,300,000 miles long, and 
stuck out over the Gulf of Mexico like a fishing-rod.  And by the 
same token any person can see that 742 years from 
now the Lower Mississippi will be only a mile and three quarters 
long, and Cairo [Illinois] and New Orleans will have joined their streets together,
and be plodding comfortably along under a single mayor and a 
mutual board of aldermen.  There is something fascinating about 
science.  One gets such wholesale returns of conjecture out of 
such a trifling investment of fact.

--- Mark Twain (1883), in {\it Life on the Mississippi}
\end{quote}

Use of  21cm data to determine the ISM 
toward an extragalactic object often requires extrapolation over 
orders of magnitude in solid angle:  from 
the area covered by the radio antenna beam, 
to the often infinitesimal area of the object under study. 
The highest angular resolution typically obtainable for a Galactic 
21cm HI emission spectrum (with good sensitivity) is $\sim 1'$, 
and this is very much larger than the size of an AGN.  
Hence the need for extrapolation.

This is a vexing subject which causes some quite respectable scientists 
 to loose their heads (I won't give explicit references, you 
know who you are).  A few take the situation 
as license to rearrange the ISM into whatever preposterous 
filigree of structure   simplifies their work --- 
contriving an ISM which appears smooth on scales constrained 
by the data, but which goes crazy in finer detail. They ignore 
the  wealth of data on the real small-scale structure 
of interstellar hydrogen.  The situation has been further confused by the 
reported discovery of an anomalous population of 
tiny, dense HI clouds,  whose significance, not to say reality,
 is now known to have been exaggerated.   Twain's warning 
against thoughtless extrapolation is especially appropriate to this 
topic.  

  {\it The Galactic ISM is not a free parameter!} 
Is there structure in the total Galactic $N_{HI}$  on all angular scales?
Yes!  Does this introduce errors  when 
extrapolating to small angles? Yes.  Are the errors important? 
Usually not!  From point to point across the sky HI 
clouds come and go,   and line components 
shift shape and velocity, but the dominant changes in 
{\it total}  $N_{HI}$ are usually 
on the largest linear scales, and do not cause  
 large fractional fluctuations over small angles ($\leq 0.5^{\circ}$).

Most structure in interstellar HI results from turbulence, characterized 
by a power-law spectrum with an exponent always less than $ -2$; 
this has been determined experimentally and 
is understood theoretically (e.g., Green 1993; Lazarian \& 
Pogosyan 2000; Deshpande, Dwarakanath, \& Goss 2000; 
Dickey et al. 2001 and references therein). 
Small angles in nearby gas (e.g., at high Galactic latitude) 
correspond to small linear scales where there is 
relatively little structure.  If the sight line intersects a 
distant high-velocity cloud then small angles may correspond to 
large spatial scales  and the  variations in that spectral component 
will be larger, but this is usually a problem for the 
{\it total} $N_{HI}$ only in directions  dominated 
by distant gas (see $\S 6$).  Cold HI may have more  
structure than warm HI, and molecular clouds even more, but 
as a practical matter, the extrapolation 
to small angles introduces large errors only when a significant part of the 
gas in a particular direction  is molecular or  of anomalous origin, 
e.g., comes from a high-velocity cloud.  Examination of 21cm 
spectra  around the position of interest should give 
adequate warning of possible structure which then would require 
higher resolution observations to measure.

A lack of appreciation for the effects of power-law turbulence 
at small angles  was one reason why 
 the early, high-resolution, VLBI studies were interpreted as 
 evidence for an anomalous  population of 
extremely dense  HI clouds with sizes of  tens of AU.  This in turn led 
 some to assume that there must be extreme fluctuations in the 
{\it total} $N_{HI}$ on very small angular scales.  Somewhere 
Mark Train was chuckling.  But 
with more complete observations (Faison 2002, Johnston et al. 2003) 
and a better understanding of how to interpret them (Deshpande 2000)
the anomaly has disappeared almost entirely.  The measured 
 small-scale fluctuations in $N_{HI}$ are likely to be entirely 
consistent with the known power laws
(Deshpande 2000).  Recent observations of  21cm absorption toward 
pulsars  probing linear scales of 0.005--25 AU find
no evidence of spatial structure at the 
$1 \sigma$ level of $\Delta e^{-\tau} = 0.035$ 
(Minter, Balser \& Karlteltepe 2003), and other 
careful observations toward pulsars suggest that some of the initial 
claims of small-angle fluctuations in HI absorption 
might be in error (Stanimirovic et al. 2003).  
The issue of small-scale structure in the ISM is 
interesting, and there are  anomalous 
directions, e.g.,  toward 3C 138, where clumping in cold gas appears 
to be significant (Faison 2002), though here the total $N_{HI}$ 
is $> 2 \times 10^{21}$ cm$^{-2}$ and not a typical extragalactic
sight-line.  

Recently Barnes \& Nulsen (2003)  
combined interferometric and single-dish data to measure 
21cm emission toward three high-latitude clusters of galaxies 
and found  limits on fluctuations 
in $N_{HI}$ on scales of $1'$--$10'$ of  $<3\%$ to $<9\%$ ($1\sigma$). 
DL90 had  suggested  that  on these angular scales 
$\sigma(N_{HI})/\langle N_{HI} \rangle \leq 10\%$, 
an estimate which has been controversial, but 
now appears somewhat conservative.  
I believe that, as concluded in DL90, 
 directions without significant $H_2$, 
and without significant anomalous-velocity HI, 
are unlikely to contain small-scale angular  
structure in HI that is  a major source of error 
in estimates of the effects of the Galaxy on extragalactic 
observations.

\section{Molecular Hydrogen and Helium}

\paragraph{Molecular Hydrogen}
  The 21cm data can be used to predict the likelihood 
that there is molecular 
gas along the line of sight, because there is usually 
 cool HI associated with molecular clouds (though the converse
 may not necessarily be true, see Gibson 2002).   
Direct observations of $H_2$ show   that $\geq10\%$ of the neutral ISM is in 
molecular form when  a sight line has a 
 reddening $E(B-V)\geq0.1$, equivalent to $N_{H} = 5.8 \times 10^{20}$
(Bohlin, Savage \& Drake 1978; Rachford et al. 2003).   This line is marked 
in Figure 2.   About half of the directions in 
the Fig.~2  sample lie above the 
$E(B-V)=0.1 $ line.  This suggests that for 
 $T_{pk} \geq 20$ K there may be molecular gas 
somewhere along the path, an    unfortunate circumstance, for  an 
accurate $N_{H_2}$ will then be  quite difficult to obtain in the 
absence of bright UV targets in these directions.  

\paragraph{Helium}

Interstellar $H_e^0$ and $H_e^+$ 
 both contribute to the opacity at $E > 13.6 $ eV 
but their abundance cannot be determined by simply 
 scaling  $N_{HI} $ and $N_{H_2}$, for a large fraction of the ISM is 
ionized (Reynolds et al. 1999).  Near the Sun, the mass 
  in ionized gas is about one-third the mass of HI,  
with substantial variations in different directions (Reynolds 1989).
The fractional He ionization in the medium seems low 
 (Reynolds \& Tufte 1995).  
Maps of the sky in $H_{\alpha}$ show the 
location of the brighter ionized regions, but 
the $H_{\alpha}$ intensity is proportional to $n_p n_e$, not 
to $N_{He}$.   The dispersion measure of pulsars  gives 
 $N_e$ exactly, but there are not enough pulsars to map 
out the Galactic $N_e$ to sufficient precision.  
 Kappes, Kerp and Richter (2003) 
studied the X-ray absorbing properties of a large area of the 
high-latitude sky and conclude that 20--50\% of the X-ray absorbing 
material is ionized and not traced by HI (see also Boulanger et al.  
2001 and references therein).  
Unlike molecular gas, whose effects can be neglected 
 in directions of low $N_{HI}$, the ionized component appears to 
cover the sky. Because we know so little about the detailed 
structure of the ionized component of the ISM,  {\it it probably 
contributes the most significant uncertainties in our understanding of 
Galactic foregrounds at high Galactic latitude.}

\begin{figure}[t]
{\includegraphics[height=0.6\vsize,width=3.0in,angle=-90]{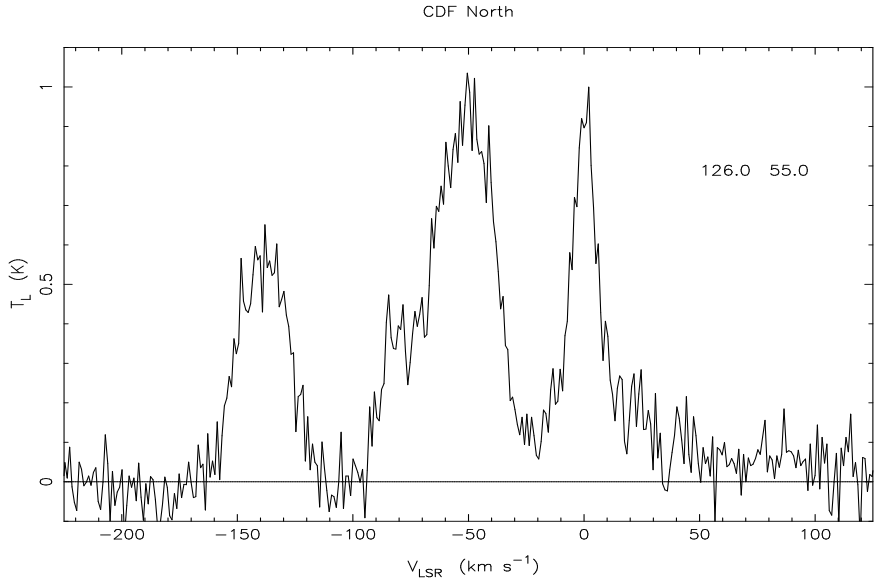}}
\caption{The 21cm spectrum toward the HDFN/CDFN at $35'$ resolution 
from the LD survey. 
}
\end{figure}

\begin{figure}[h]
{\includegraphics[height=3.5in,width=3.5in,angle=-0]{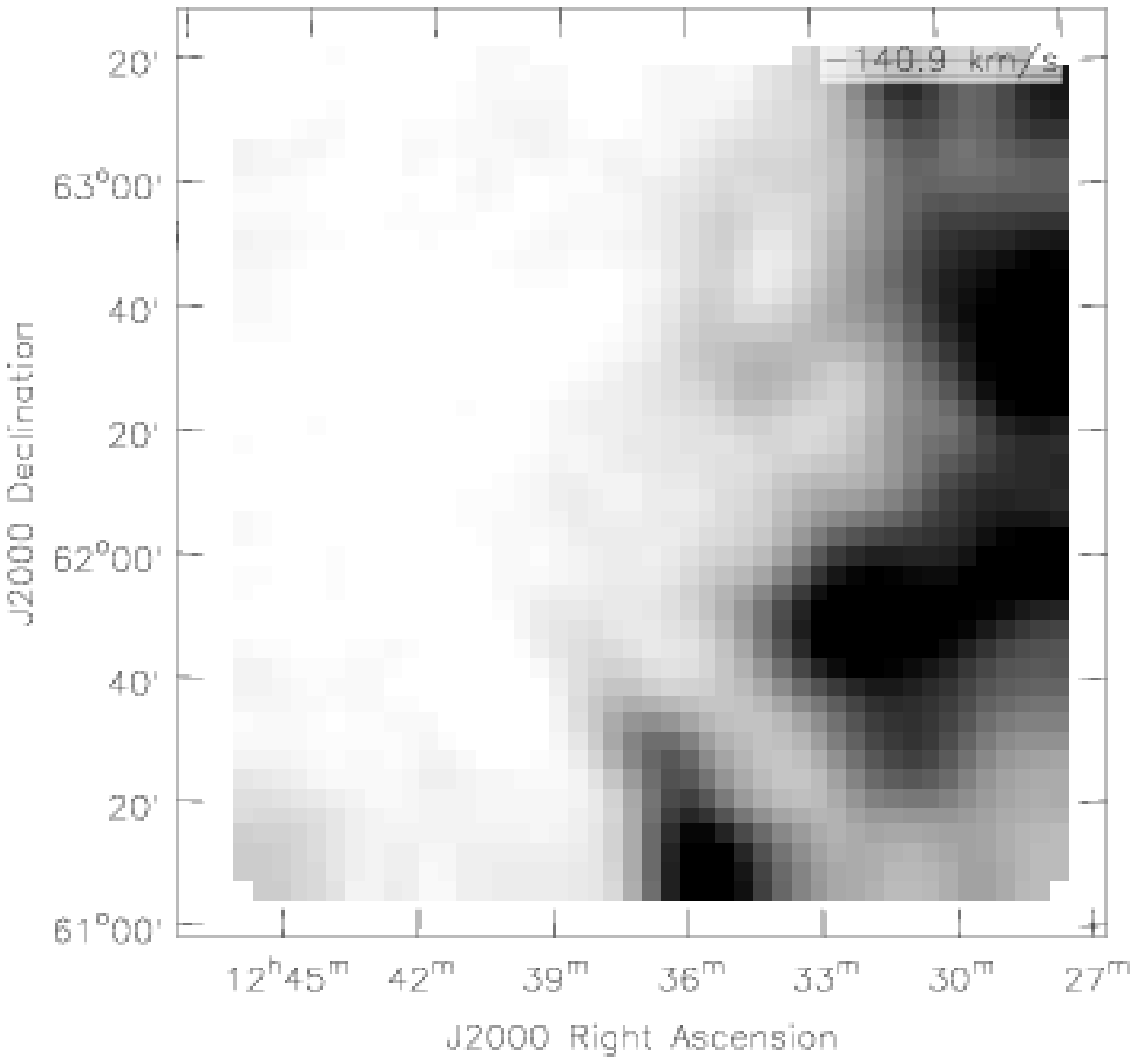}}
{\caption{The structure in the high-velocity cloud at $-140.9 $ 
km s$^{-1}$ toward 
the Chandra Deep Field North  as measured 
 at $9'$ resolution with the Green Bank Telescope (Lockman 2004, 
in preparation).  The emission line  varies from $T_{pk}>2$ K to 
$T_{pk}<0.05$ K across the field.  The CDFN is at 
 $12^h36^m50^s +62^{\circ}13'00''$ }}
\end{figure}

\section{ The Chandra Deep Field North (CDFN)}

The Chandra Deep Field North is an example of a direction 
which lacks dense gas, and thus one set of complexities, but has other 
features of interest (this direction is also the Hubble Deep Field, but 
as my comments are probably of more interest to X-ray than optical 
astronomers, I will keep the X-ray name).  
Figure 3 shows a 21cm spectrum from the LD survey at 
$35'$ resolution toward the CDFN ($\ell,b=126^{\circ}+55^{\circ}$). 
As a Galactic astronomer I find this sight-line fascinating 
because it intersects  a  high-velocity cloud,
an intermediate velocity cloud, and low-velocity ``disk'' gas, 
(from left to right in the spectrum), 
containing, respectively, 20\%, 50\% and 30\% of the total $N_{HI}$. 
This gas is almost certainly optically thin in the 21cm line. 
The low-velocity gas further shows evidence of two components, 
one warm and one cool.  
It is uncommon to find a sight line where anomalous-velocity 
gas dominates the total $N_{HI}$, but that is the case for the 
CDFN and nearby areas of the sky.  

Observations of the CDFN with the 100 meter Green Bank Telescope 
(GBT) of the NRAO at $9'$ resolution 
show a spectrum similar to Figure 3 at the low and intermediate 
velocities, but only $1/3$ as bright toward the high-velocity cloud.  
Figure 4 shows why: the high-velocity cloud 
 has a gradient of nearly two orders of magnitude 
in $N_{HI}$ near the CDFN.   It becomes the brightest component in the 
entire spectrum to the west, 
 contributing nearly $10^{20}$ cm$^{-2}$ or more than $40\%$ of
the total column density, while to the east it is almost undetectable. 
This high degree of angular structure 
is typical of high-velocity clouds, which lie far beyond the 
local disk gas (Wakker \& van Woerden 1997).  The intermediate 
velocity gas over this field has a smaller dependence on angle,
with a factor $\sim 2$ change in $N_{HI}$ and  
most structure to the south and east.

In a direction (like the CDFN) where much of the hydrogen is in 
 a high-velocity cloud, the 
 total  $N_{HI}$ is likely to be a poor predictor of many interesting 
quantities like $S_{100\mu}$ and $E(B-V)$.  
For example, high-velocity clouds have 
a lower emissivity per H atom at $100\mu$ than disk  and 
intermediate-velocity gas  because they lack of dust and/or 
have low heating (Wakker \& Boulanger 1986).  
The particular high-velocity cloud that crosses the CDFN 
has an  abundance of metals only about $10\%$ that of 
solar (Richter et al. 2001).  The intermediate velocity HI component 
also probably has different properties than disk gas at some wavelengths 
because of its different dynamical history.  
Thus, unlike most directions on the sky, where the 
 kinematics of Galactic HI is irrelevant to its
effects on extragalactic observations, the direction of the CDFN may be an 
exception, and require a specific HI component analysis.  
In contrast to the CDFN, preliminary GBT observations of 
the Chandra Deep Field South show that it has a HI 
 spectrum dominated by a single, weak, low-velocity line.  Quite boring 
compared to the CDFN.

\section{Concluding Comments}
The 21cm line can be a powerful tool for estimating the influence 
of the Galactic ISM on extragalactic observations, but it must be 
used with some thought, and it does not give the complete 
picture.  Unresolved angular structure in the 21cm data is unlikely 
to dominate the error budget in most directions.  
For $N_{HI} > 6 \times 10^{20}$ cm$^{-2}$  it is probable 
that there is some  molecular hydrogen  along the sight line and 
 the total $N_H$ will be difficult to determine.  But in my opinion, 
it is the poorly-understood ionized component of the ISM 
which introduces the most serious 
uncertainties for directions with little molecular gas,  
and it affects observations in all directions. 
We are entering an era when highly accurate 21cm data will be available 
over the entire sky, and then, (though likely even now) the limits on 
understanding the influence of the ISM on extragalactic observations 
will lie not in the uncertainties in $N_{HI}$, 
but in $N_{H_2}$ and $N_{He}$, and in the relationship between 
the dust and the gas.  Every observation of an extragalactic object is 
an opportunity to learn something about the ISM.  We should not let 
such opportunities go to waste. 

\begin{acknowledgments}
I thank J.M. Dickey, M. Elvis, C.E. Heiles, A.H. Minter, and R.J. Reynolds for 
comments on the manuscript, and D. McCammon for supplying Figure 1.  

\end{acknowledgments}

\end{document}